\documentclass[aps,preprint,showpacs,amsmath,amssymb,groupedaddress,floatfix]{revtex4}
\usepackage{graphicx}
\usepackage{dcolumn}
\usepackage{bm}
\usepackage{amsmath}
\usepackage{amssymb}
\usepackage[usenames]{color}

\usepackage[T1]{fontenc}

\begin{document}

\title{Large-scale calculations of supernova neutrino-induced reactions in Z=8-82 target nuclei}
\author{N. Paar$^1$}
\email{npaar@phy.hr}
\author{H. Tutman$^1$}
\author{T. Marketin$^{1,3}$}
\author{T. Fischer$^{2,3}$}
\affiliation{$^1$Physics Department, Faculty of Science, University of Zagreb, Croatia}
\affiliation{$^2$GSI Helmholtzzentrum f\"{u}r Schwerionenforschung, Planckstra\ss e 1, D-64291 Darmstadt, Germany}
\affiliation{$^3$Institut f\"ur Kernhysik, Technische Universit\"at Darmstadt,
Magdalenenstra{\ss}e 12, 64289 Darmstadt, Germany}
\date{\today}

\begin{abstract}
{\bf Background:}
In the environment of high neutrino-fluxes provided 
in core-collapse supernovae or neutron star mergers, neutrino-induced reactions with nuclei contribute to the nucleosynthesis processes. A number of terrestrial neutrino detectors are based on inelastic neutrino-nucleus scattering and modeling of the respective cross sections allow predictions of the expected detector reaction rates.
{\bf Purpose:}
To provide a self-consistent microscopic description of neutrino-nucleus cross sections involving a large pool of $Z = 8 - 82$ nuclei for the implementation in models of nucleosynthesis and neutrino detector simulations.
{\bf Methods:}
Self-consistent theory framework based on relativistic nuclear energy density functional is employed to determine the nuclear structure of the initial state and relevant transitions to excited states induced by neutrinos. The weak neutrino-nucleus interaction is employed in the current-current form and a complete set of transition operators is taken into account.
{\bf Results:}
We perform large-scale calculations of charged-current neutrino-nucleus cross sections, including those averaged over supernova neutrino fluxes, for the set of even-even target nuclei from oxygen toward lead ($Z = 8 - 82$), spanning $N = 8 - 182$ (OPb pool). The model calculations include allowed and forbidden transitions up to $J = 5$ multipoles.
{\bf Conclusions:}
The present analysis shows that the self-consistent calculations result in considerable differences in comparison to previously reported cross sections, and for a large number of target nuclei the cross sections are enhanced. Revision in modeling r-process nucleosynthesis based on a self-consistent description of neutrino-induced reactions would allow an updated insight into the origin of elements in the Universe and it would provide the estimate of uncertainties in the calculated element abundance patterns. 

\end{abstract}

\pacs{21.30.Fe, 21.60.Jz, 23.40.Bw, 25.30.-c}
\maketitle

\bigskip \bigskip

\section{Introduction}

Neutrino-induced reactions on nuclei play an important role in nuclear astropyhsics, in particular during core-collapse supernova evolution and nucleosynthesis~\cite{Bru.91,Jan.07}. Elastic neutrino scattering on nuclei and nucleons determines the neutrino trapping and the diffusion time scale of the outwards streaming neutrinos~\cite{Pin.06}. A variety of processes contribute to the energy loss in stellar interiors, e.g. pair, photo-, plasma, nucleon-nucleon bremsstrahlung, and recombination neutrino processes~\cite{Ito.96}. In addition, inelastic neutrino-electron scattering mainly thermalizes the neutrino spectra~\cite{Mez.93} and inelastic neutral-current neutrino-nucleus scattering has a large contribution to the opacity~\cite{Ger.75,Ger.75b}, during the core collapse and subsequent explosion phase. Additionally, in the environment of exploding massive stars, which has long been explored as a possible site for the r-process nucleosynthesis, charged-current neutrino-nucleus reactions play an important role in the production of chemical elements.
As pointed out in Ref.~\cite{Hek.00}, there are many interesting effects of neutrino-induced reactions before, during, and after the r-process. To explore their role in nucleosynthesis reliable neutrino-nucleus reaction rate compilations involving also nuclei with large neutron excess are required. Available data on neutrino-nucleus cross sections are limited to deuterium~\cite{Wil.80}, and  $^{12}$C and $^{56}$Fe target nuclei, obtained by the LSND~\cite{Ath.97} and KARMEN~\cite{Bod.94,Mas.98} collaborations, and at LAMPF~\cite{Kra.92}. Therefore, only theoretical approaches can provide cross sections for a large number of target nuclei that are involved in various applications of neutrino physics and astrophysics. Modeling neutrino-nucleus interactions is also important in view of the current research and development of neutrino detectors, e.g., for supernova and solar neutrinos, neutrinos produced in laboratories, and geoneutrinos. The ongoing and planned neutrino detector facilities involve a variety of target materials, induced reactions and scientific objectives, e.g., MOON~\cite{Eji.08}, MiniBooNE~\cite{Agu.09}, MINOS~\cite{Ada.06}, SNO+~\cite{Chr.09}, OPERA~\cite{Aga.10}, LVD (Large Volume Detector)\cite{Aga.07}, ORLaND experiment proposal at the Spallation Neutron Source (SNS)~\cite{Efr.05}, NOvA neutrino experiment~\cite{Har.05}, Daya Bay reactor neutrino experiment~\cite{An.12}, etc. There is also interesting concept of beta-beams for the production of neutrinos by using $\beta$-decay of boosted radioactive ions~\cite{Zuc.02,Vol.04}.  The proposal to establish a beta-beam facility that could produce low-energy neutrino beams in the 100 MeV energy range would allow direct insight into
the neutrino-induced reactions in nuclei as well as its underlying structure involving a non-trivial combination of nuclear allowed and forbidden transitions~\cite{Vol.04}.

At low neutrino energies, the neutrino-induced reactions are sensitive to the properties of nuclei involved, i.e. their initial and excited states. Therefore it is necessary to employ a microscopic framework providing reasonable description of nuclear structure properties. Over the past years, a variety of advanced microscopic models have been developed and employed in studies of charged-current neutrino-induced reactions at low energies, also including various particle decay channels. In particular, these include the nuclear shell model~\cite{Hax.87,Eng.96,Hay.00,Sam.02,Yos.08,Suz.09,Suz.11}, random phase approximation (RPA)~\cite{Hek.00,Aue.97,Sin.98,Vol.00,Vol.02,Eng.03,Ath.06}, continuum RPA (CRPA)~\cite{Kol.92,Kol.95,Jac.99,Jac.02,Bot.05}, and the hybrid model which combines the shell model for allowed transitions, with the RPA to account for the forbidden transitions~\cite{Kol.99,Toi.01,Kol.03,Suz.09}. Frameworks based on quasiparticle RPA (QRPA) have also recently been developed, based on Skyrme functionals~\cite{Vol.00,Laz.07}, Brueckner G matrix employed for two-body interaction by solving the Bethe-Salpeter equation using Bonn-CD potential~\cite{Che.10,Che2.10,Che.11}, and projected QRPA~\cite{Krm.05,Sam.08,Sam.11}.  The Fermi gas model~\cite{Wal.75,Gai.86,Kur.90} and Fermi liquid theory (FLT)~\cite{Lyk.07} have also been employed in studies of low-energy charged-current neutrino-nucleus reactions. At finite temperature in stellar environments, thermal population of the excited states may enhance the weak interaction rates and cross sections 
at low neutrino energies~\cite{Dzh.10,Dzh.11}.

Despite considerable progress in the development of advanced theoretical frameworks, up to this time only a limited number of microscopic models have been employed in large-scale calculations of neutrino-induced reactions and their implementation in supernova simulations. Because the complete modeling of neutrino-induced reactions necessitates the inclusion not only of Gamow-Teller transitions but also contributions from forbidden transitions and other higher multipoles, covering a large pool of target nuclei represents a computationally very demanding problem. Although the shell model provides a very accurate description of ground state wave functions, the description of high-lying excitations necessitates the use of large model spaces which often leads to computational difficulties, making the approach applicable essentially only to allowed transitions in light and medium-mass nuclei. For systematic studies of neutrino-nucleus cross sections throughout the nuclide chart including the heavy nuclei, microscopic calculations must therefore be performed using models based on the QRPA. The first global calculations of electron neutrino-nucleus cross sections have been conducted using the gross theory of $\beta$-decay~\cite{Ful.95,McL.95}. Two microscopic frameworks have been employed in large-scale calculations of neutrino-induced reactions i) extended Thomas-Fermi plus Strutinsky integral (ETFSI) and continuum quasiparticle random phase approximation (CQRPA)~\cite{Bor.00}, and ii) RPA with Landau-Migdal (LM) force, using Wood-Saxon potential (WS) to determine the single-particle basis of target nuclei~\cite{Lan.01}. There are no up-to-date systematic calculations conducted in the framework of energy density functional. 

Recently the framework based on relativistic nuclear energy density functional (RNEDF) has been introduced in modeling the charged-current neutrino-nucleus cross sections~\cite{Paa.08}, and it has also been extended for applications in modeling neutral-current neutrino-induced reactions~\cite{Djapo.12}. In the case of iron group nuclei, comparison of the results for charge exchange reactions obtained using Skyrme functionals and the RNEDF, as well as with the shell model, showed reasonable theoretical uncertainty inherent in modeling neutrino-nucleus cross sections. However, in view of applications in supernova and r-process simulations, which also involve unstable nuclei far from the valley of stability, where no experimental data are available, it is necessary to provide independent insight into relevant neutrino-induced processes from various models and effective interactions. The main objective of this work is conducting large-scale calculations of charged-current neutrino-nucleus cross sections in a large pool of nuclei from oxygen towards lead, based on the RNEDF. 
In addition to the overall cross sections covering the range of neutrino energies up to 100 MeV, calculations also include the cross sections averaged over neutrino fluxes for the range of temperatures characteristic of various stages of stellar evolution. In addition to the Fermi-Dirac neutrino spectra we also apply supernova spectra from a recent simulation that includes three-flavor Boltzmann neutrino transport. Particular aim of this work is to emphasize the role of forbidden transitions in modeling neutrino-nucleus cross sections in large pool of nuclei, including both stable and nuclei away from the valley of stability. The relevance of forbidden transitions has already been discussed in several
studies, e.g. Refs.~\cite{Sur.98,Vol.04,Laz.07}.  In the study of neutrino capture by r-process waiting point nuclei, first forbidden strength, together with the low-lying Gamow-Teller transitions, increased 
the rate of neutrino scattering from very neutron-rich nuclei by a factor of at least 2 and in some 
instances even by a factor of 5~\cite{Sur.98}. As pointed out in microscopic calculations based on
Skyrme functionals in Ref.~\cite{Laz.07}, the properties of
forbidden states are closely related to the neutrino-nucleus cross sections, and could be extracted 
by using neutrinos from low-energy beta-beams.

The paper is organized as follows. Section II contains the basic theoretical background for the neutrino-nucleus cross sections in the charged-current channel based on weak Hamiltonian and the RNEDF. The results of large-scale calculations of the neutrino-nucleus cross sections are illustrated and discussed in Sec. III. The conclusions of the present work are summarized in Sec. IV.

\section{Theoretical background}

In the present work we explore the charged-current neutrino-nucleus process,
\begin{equation}
\nu_e + X_{(Z,N)}  \rightarrow X^{*}_{(Z+1,N-1)} + e^- \;,
\end{equation}
where the incoming electron neutrino $(\nu_e)$ induces charge-exchange reaction in target nucleus $X(Z,N)$. The general formalism of the neutrino-nucleus cross sections, derived assuming the Hamiltonian of the weak interaction in the current-current form, is given in details in Refs.~\cite{Con.72,Wal.75}. The cross sections include the transition matrix elements between the initial $|J_i \rangle$ and final nuclear state $|J_f \rangle$, for the charge $\hat{\mathcal{M}}_J$, longitudinal $\hat{\mathcal{L}}_J$, transverse electric $ \hat{\mathcal{T}}_J^{EL}$, and transverse magnetic $\hat{\mathcal{T}}_J^{MAG}$ multipole operators~\cite{Con.72}. In the present work the RNEDF is employed in calculations of the transition matrix elements contributing to the neutrino-nucleus cross sections. The RNEDF has already been successfully employed in studies of giant resonances and exotic modes of excitation~\cite{Vre.99,Vre.04,PVKC.07,PPP.05,Paa2.09,Kha.11}, $\beta$-decay rates of r-process nuclei~\cite{Nik.05}, muon capture~\cite{Mar.09} and stellar electron capture rates~\cite{Niu.11}, and in constraining the neutron skin in nuclei~\cite{Vre.03,Kli.07}.  More details about the implementation of the RNEDF in modeling charged-current neutrino-nucleus reactions are given in Refs.~\cite{Paa.08,Paa.11}. 

The RNEDF based framework employs the self-consistent mean field for nucleons and minimal set of meson fields; isoscalar scalar $\sigma$-meson $(J^{\pi}=0^+, T=0)$, isoscalar vector $\omega$-meson  $(J^{\pi}=1^-, T=0)$ and the isovector vector $\rho$-meson $(J^{\pi}=1^-, T=1)$, supplemented with the electromagnetic field. The meson-nucleon interaction is included with a minimal set of the interaction terms, where the vertex functionals include explicit dependence on the vector density. 
The nuclear ground state properties are described using the relativistic Hartree-Bogoliubov model (RHB), and relevant transitions induced by neutrinos are calculated in the relativistic quasiparticle random phase approximation (RQRPA). More details on the RHB model based on effective density-dependent interactions are given in Ref.~\cite{Nik.02}. The RQRPA is formulated in the canonical single-nucleon basis of the RHB model~\cite{Paa.03,NVR.02}. In modeling the neutrino-nucleus cross sections, important advantage of this framework is that it is fully consistent in view of the effective interactions employed. In the particle-hole $(ph)$ and pairing $(pp)$ channels, the same interactions are used in the RHB equations that determine the canonical quasiparticle basis, and in the matrix equations of the RQRPA. In this way, one can employ the same RNEDF in description of the weak processes throughout the nuclide map without any additional adjustments to the properties of specific target nuclei under consideration. For the model parameters that determine the density-dependent couplings and the meson masses, in this work DD-ME2 parameterization is used~\cite{Lal.05}. The pairing correlations in open shell nuclei are described by the finite range Gogny interaction, with parameterization D1S~\cite{Ber.91}.

Complete calculation of inelastic neutrino-nucleus reactions spanning the range of neutrino energies up to $\approx$ 100 MeV necessitates the inclusion of a number of transitions with various multipoles $J$~\cite{Paa.08}. Although higher-order multipoles have rather small contributions at low incoming neutrino energies, these can not be neglected at energies about tens of MeV. In the present study, multipoles up to $J=5$ have been taken into account. The large scale calculations of neutrino-nucleus cross sections, involving more than 1000 nuclei and a complete set of all multipoles up to $J=5$ with both parities, necessitate considerable computational effort. Therefore, for the purposes of the present work, computational framework has been developed using parallel computing methods based on Message Passing Interface (MPI)~\cite{Gro.94} for the implementation on cluster and grid computer systems. 

\section{Results and discussion}
\begin{figure}
%\vspace*{1cm}
\centering
\includegraphics[scale=0.5]{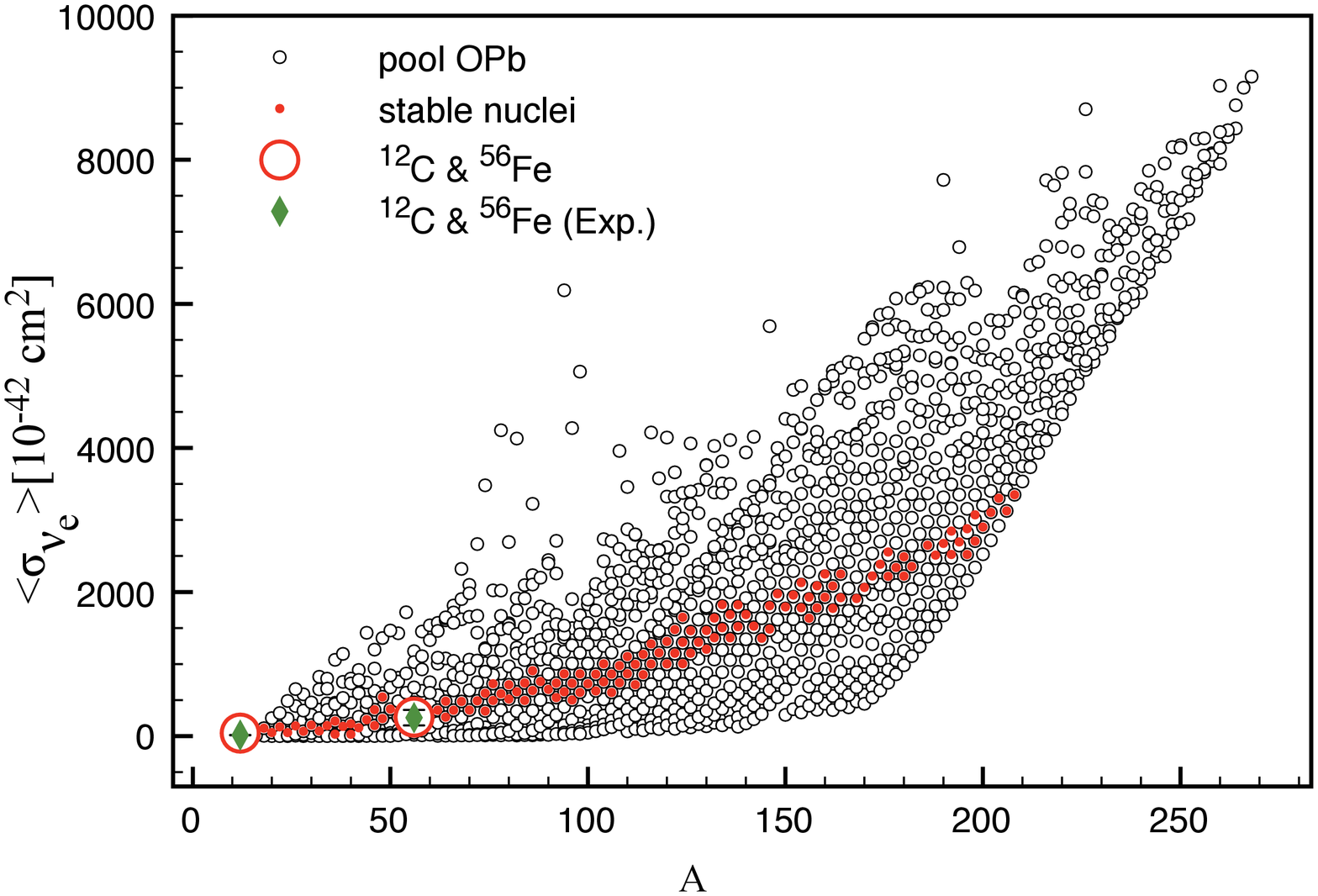}
\caption{(Color online) Inclusive neutrino-nucleus cross sections for
target nuclei in the OPB pool, averaged over the Michel neutrino flux
from muon DAR. The cross sections for stable nuclei are denoted by
filled circles. The emphasized results for $^{12}$C and $^{56}$Fe are shown in
comparison to the experimental data (KARMEN)~\protect\cite{Bod.94,Mas.98}.}
\label{all_isotopes_michel.eps}
\end{figure}

By employing the model outlined in the previous section, we have conducted large-scale calculations of  $(\nu_e,e^-)$ reactions in the OPb pool of even-even target nuclei spanning the range from oxygen toward lead (Z=8-82) with neutron number N=8-182. The calculations include excitations of all multipoles up to $J=5$ and both parities. In the following we explore the systematic behavior of the overall cross sections throughout the pool under consideration, including the cross sections averaged over the Michel neutrino flux obtained from the decay at rest (DAR) of $\mu^{+}$~\cite{Kra.92}
\begin{equation} \label{eq:Michel}
f_{\text{M}}(E_{\nu}) = \frac{96 E_{\nu}^{2}}{m_{\mu}^{4}} \left( m_{\mu} - 2 E_{\nu} \right).
\end{equation}
To simulate the supernova neutrino spectrum, at first instance we use the Fermi-Dirac distribution
\begin{equation}
f_\text{FD}(E_{\nu}) =  \frac{1}{T^3}  \frac{E_{\nu}^2} {\exp \left[ (E_{\nu}/T)-\eta \right ] + 1  } \; .
\label{fermidirac}
\end{equation}
In Fig.~\ref{all_isotopes_michel.eps} the inclusive neutrino-nucleus cross sections, averaged over the Michel neutrino flux, are shown for the OPb pool of nuclei. Stable nuclei in the pool are denoted by filled circles. In addition, the calculated cross sections for $^{12}$C and $^{56}$Fe are especially emphasized in comparison to the KARMEN experimental data~\cite{Bod.94,Mas.98}. One can observe that the pool of nuclei calculated with the same energy density functional without any adjustments of the model parameters fits nicely into two experimental data points. The cross sections increase systematically with increasing neutron number, resulting in values larger up to a factor $\approx 2-3$ in comparison to those in the valley of stability. We note that sharp edge of the data at large $A$ denotes boundary values for the pool under consideration, obtained for Pb isotope chain.

\begin{figure}
%\vspace*{1cm}
\centering
\includegraphics[scale=0.5]{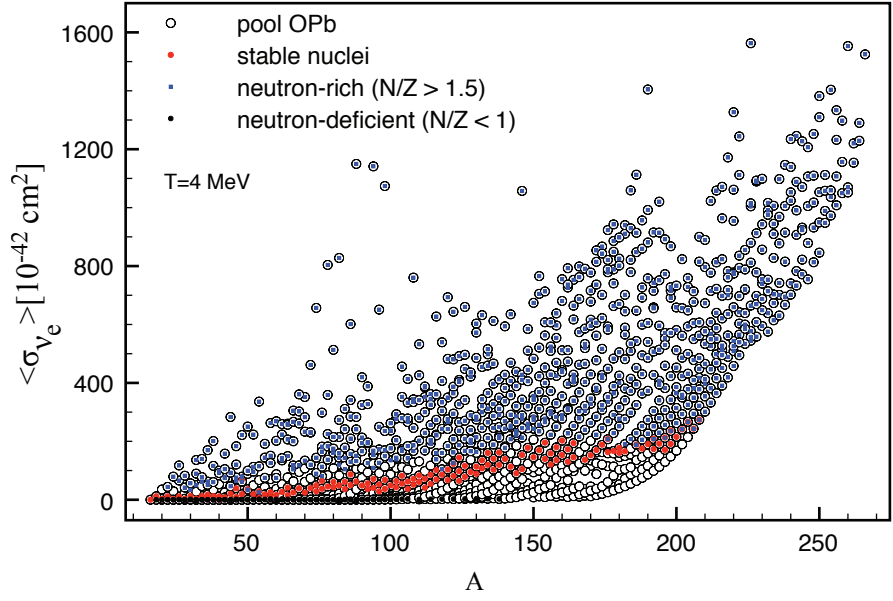}
\caption{(Color online) Inclusive neutrino-nucleus cross sections for
target nuclei in the OPb pool averaged over the Fermi-Dirac distribution 
for $T = 4$ MeV, $\eta = 0$. The cross sections for various groups of target
nuclei are emphasized: stable nuclei (filled red circles), neutron-deficient nuclei with $N/Z < 1$ (filled black circles), neutron-rich nuclei with $N/Z > 1.5$ (solid squares).}
\label{all_isotopes_T4.eps}
\end{figure}

\begin{figure}
%\vspace*{1cm}
\centering
\includegraphics[scale=0.4]{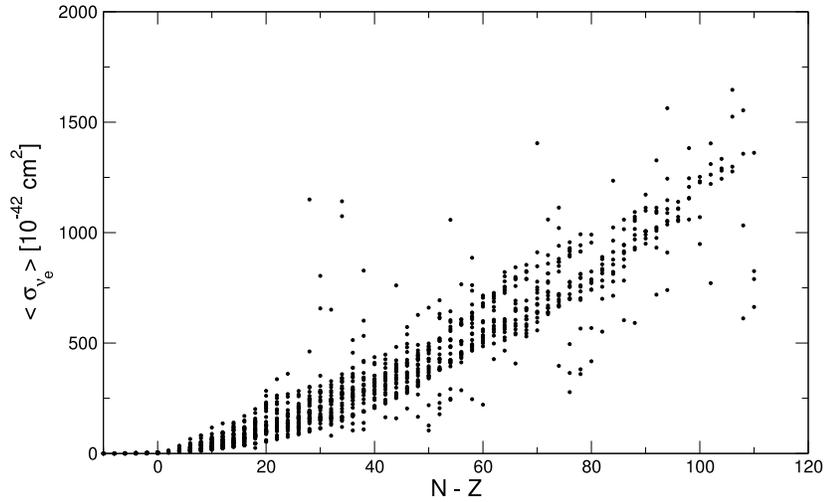}
\caption{Inclusive neutrino-nucleus cross sections averaged over the Fermi-Dirac distribution
with $T = 4$ MeV is plotted versus the difference between the neutron and the proton number.}
\label{fig:nzdiff}
\end{figure}

\begin{figure}
%\vspace*{1cm}
\centering
\includegraphics[scale=0.5]{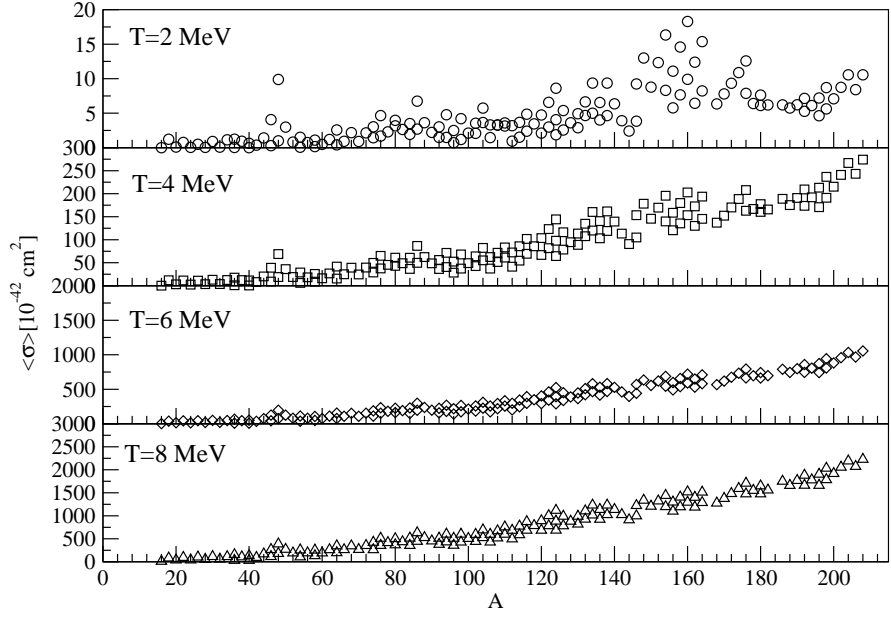}
\caption{Inclusive neutrino-nucleus cross sections for
stable target nuclei averaged over the Fermi-Dirac distribution 
for $T = 2-8$ MeV, $\eta = 0$.}
\label{neutrino_stable_T_2_8.eps}
\end{figure}

Figure~\ref{all_isotopes_T4.eps} shows the $(\nu_e,e^-)$ cross sections in the OPb pool of nuclei, averaged over the supernova neutrino flux given by Eq. (\ref{fermidirac}) in the case of $T = 4$ MeV and $\eta = 0$. The results can be compared to those of the stable target nuclei. While the cross sections display a similar pattern as in the previous case with the Michel spectrum, the details of the data, however, depend on $(T,\eta)$ of the neutrino spectrum. The cross sections for various groups of target nuclei are separately displayed in Fig.~\ref{all_isotopes_T4.eps}: stable nuclei, proton-rich nuclei constrained by $N/Z < 1$, and neutron-rich nuclei with with $N/Z > 1.5$. The cross sections exhibit a systematic behavior throughout the nuclide map. In the case of neutron rich nuclei, the cross sections are larger in comparison to the stable nuclei, due to the increased number of neutrons that participate in charge-exchange neutrino-induced reactions. For proton-rich nuclei, the reaction pattern is opposite due to \th{the smaller} number of neutron-proton $2qp$ configurations, i.e., the $(\nu_e,e^-)$ cross sections are considerably reduced in comparison to those of stable nuclei. 
This point is further illustrated in Fig.~\ref{fig:nzdiff} where we plot the same cross sections versus the difference between the neutron and the proton numbers. For nuclei with $N - Z > 1$ the cross sections increase significantly with increasing number of excess neutrons. Although some scattering is apparent in the data, most of the results cluster along an almost linear function of the number of excess neutrons. Temperature dependence of neutrino-induced reactions within the OPb pool is illustrated in Fig.~\ref{neutrino_stable_T_2_8.eps}, where the cross sections averaged over the Fermi-Dirac distribution ($\eta = 0$) are shown for stable target nuclei in the range $T = 2 - 8$ MeV. At low temperatures, the cross sections appear rather scattered due to the stronger dependence on low-energy excitations in nuclei. However, for higher T, the cross sections are rather smooth and considerably larger due to the inclusion of a number of multipoles contributing at higher neutrino energies.
The full set of calculations has been completed for the OPb pool in the range of temperatures $T = 0 - 10$ MeV and complete tables are available on request.

\begin{figure}
%\vspace*{1cm}
\centering
\includegraphics[scale=0.5]{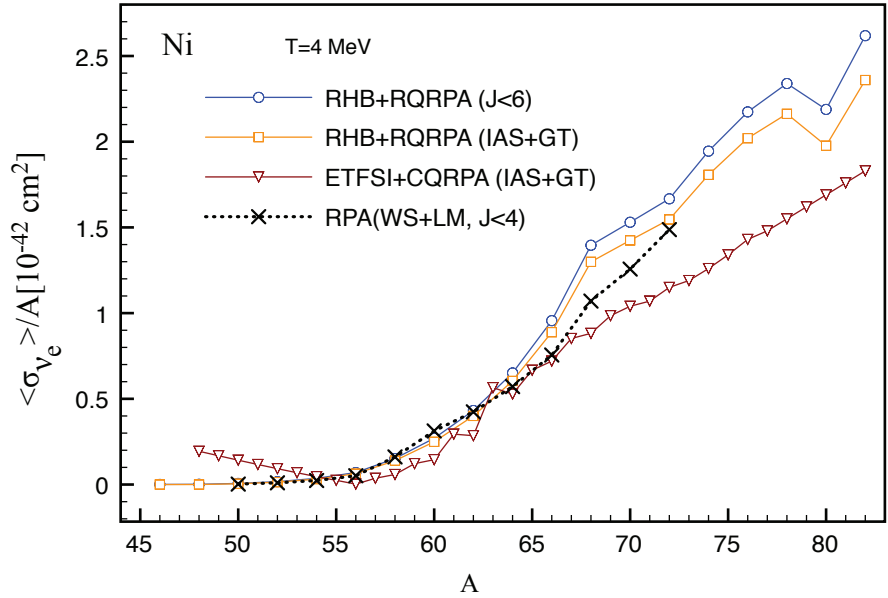}
\caption{(Color online) Flux averaged supernova $\nu_e$-nucleus cross sections for Ni isotopes
($T = 4$ MeV, $\eta = 0$). The RHB+RQRPA cross sections calculated using
all $J\leq5$ transitions are compared to those of ETFSI+CQRPA which include IAS+GT transitions ~\protect\cite{Bor.00}
and RPA (WS+LM) which cover multipoles up to J=3~\protect\cite{Lan.01}.}
\label{csm_Ni_isotopes_T4.eps}
\end{figure}

\begin{figure}
%\vspace*{1cm}
\centering
\includegraphics[scale=0.5]{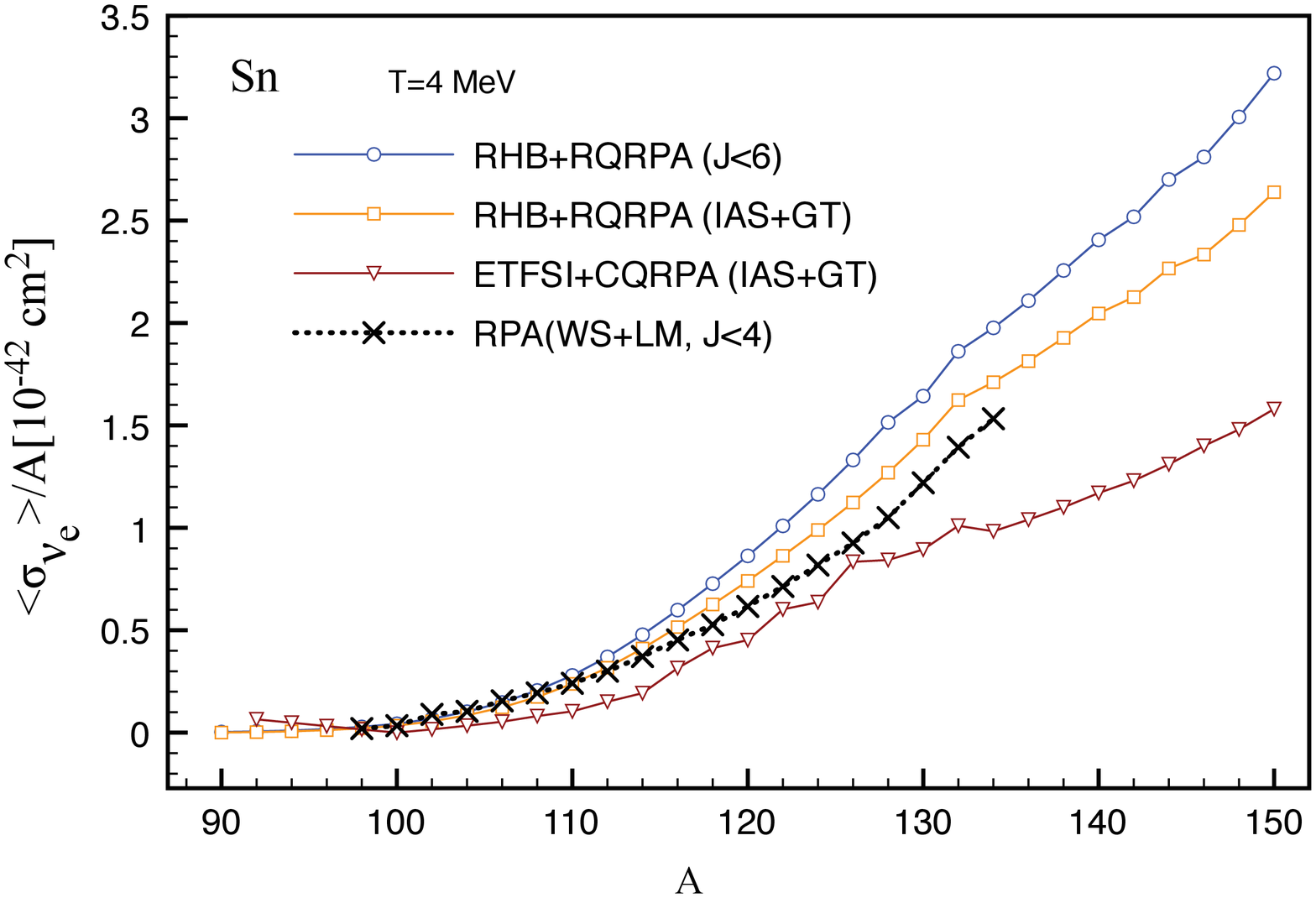}
\caption{(Color online) The same as in Fig.~\protect\ref{csm_Ni_isotopes_T4.eps}, but for Sn isotopic chain.}
\label{csm_Sn_isotopes_T4.eps}
\end{figure}

\begin{figure}
%\vspace*{1cm}
\centering
\includegraphics[scale=0.5]{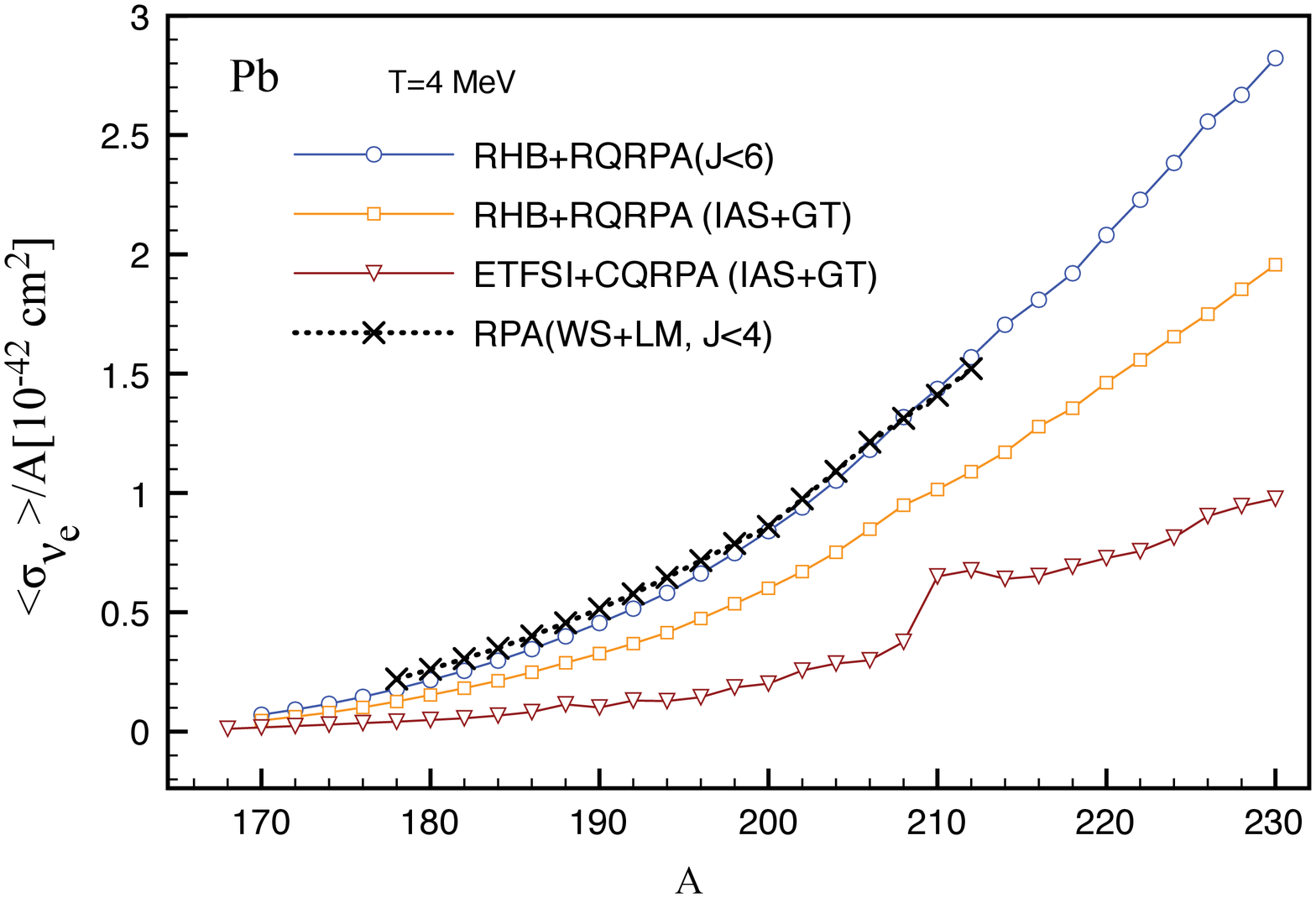}
\caption{(Color online) The same as in Fig.~\protect\ref{csm_Ni_isotopes_T4.eps}, but for Pb isotopic chain.}
\label{csm_Pb_isotopes_T4.eps}
\end{figure}

In view of applications in astrophysical models and in predicting the detector response to neutrinos involving various target nuclei, it is crucial to assess the sensitivity of the neutrino-nucleus cross sections on the theoretical frameworks and effective interactions employed. In the case of iron group nuclei, it
has been shown that by employing different microscopic models, one can estimate reasonable theoretical uncertainty in neutrino-nucleus cross sections averaged over the Michel spectrum, i.e., for $^{56}$Fe $<\sigma>_{th} = (258 \pm 57) \times 10 ^{-42}$ cm$^2$~\cite{Paa.11}. Since for nuclei far from the valley of stability various models result in larger differences in nuclear structure properties, one could also expect larger sensitivity in the results of modeling neutrino-induced reactions. In the present analysis theoretical uncertainties are assessed on the basis of the OPb pool of nuclei. Figures ~\ref{csm_Ni_isotopes_T4.eps}, ~\ref{csm_Sn_isotopes_T4.eps} and ~\ref{csm_Pb_isotopes_T4.eps} show the dependence of the flux-averaged neutrino-nucleus cross sections per nucleon ($T = 4$ MeV, $\eta = 0$) on the neutron number in the cases of Ni, Sn, and Pb isotopic chains, respectively. In these figures we compare our results to the available sets of cross sections based on two other theoretical frameworks: ETFSI + CQRPA~\cite{Bor.00} and RPA (WS+LM)~\cite{Lan.01}, where the former 
contains only contributions from the Gamow-Teller and Fermi transitions, while the latter includes all transitions up to $J = 3$. The results of the present RQRPA analysis for $<\sigma_{\nu_e}>/A$ are shown separately for the full calculation including all transitions up to $J = 5$, and partial cross sections obtained only for the isobaric analog and Gamow-Teller transitions. 

The RQRPA results for $N > Z$ nuclei, obtained taking into account only the Fermi and the Gamow-Teller transitions, are consistently higher than the corresponding values calculated using the ETFSI + CQRPA framework for all three isotopic chains. The deviation is rather small in the region around the valley of stability in Ni and Sn isotopes, but increases with additional neutrons. In the case of the Pb isotopic chain, the difference is significant even for the lightest isotopes considered in this study. For nuclei with the proton-to-neutron ratio $Z/N < 1$, the ETFSI + CQRPA framework predicts a change of the general trend and an anomalous increase of the cross sections with a corresponding decrease of the number of neutrons. The agrement between results that take into accout forbidden transitions, RQRPA and RPA(WS+LM), is much better in all three isotopic chains, although for very neutron-rich Ni and Sn isotopes the RQRPA predicts higher values of the flux-averaged cross section. In the case of the Pb isotopic chain, the agreement between the RQRPA and RPA(WS+LM) cross sections is excellent for all the isotopes under consideration. There are several reasons for the shown deviations. Each model employs different effective interactions that result in variations of the excitation pattern contributing to the cross sections, and the present calculations are based on a fully self-consistent approach to the neutrino-induced reactions. The three models agree best for nuclei around the valley of stability, where most of the experimental data on nuclear structure properties are available. In the very neutron rich region the differences between various models increase and grow relatively large. However, in all three isotopic chains the agreement of the RQRPA results is much better with the values obtained using the RPA (WS + LM) model than with the ETFSI + CQRPA, partly indicating the importance of the forbidden transitions. 

\begin{figure}
%\vspace*{1cm}
\centering
\includegraphics[scale=0.5]{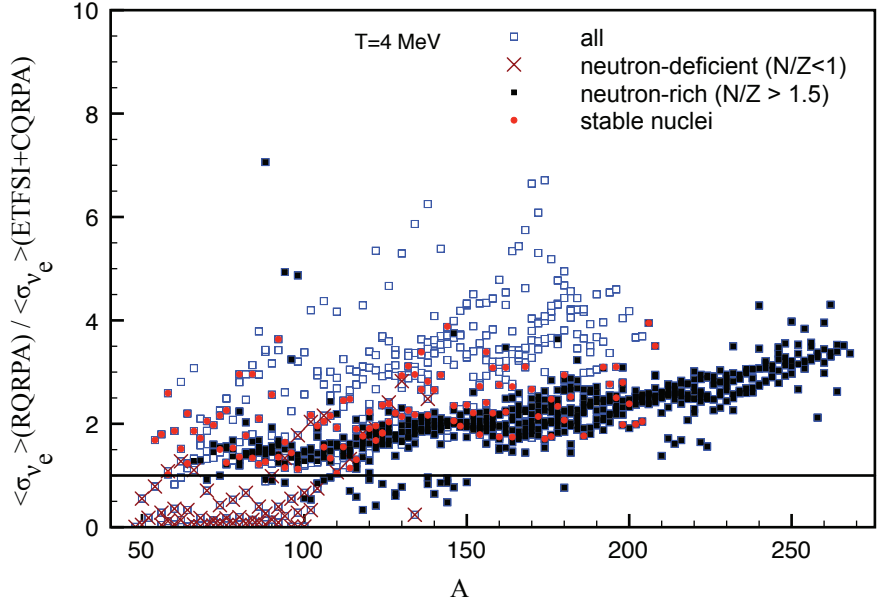}
\caption{(Color online) The ratio of the cross sections averaged over supernova neutrino flux ($T = 4$ MeV, $\eta = 0$) calculated using the RHB+RQRPA and ETFSI+CQRPA~\protect\cite{Bor.00} models. Various groups of nuclei are separately denoted: stable nuclei (solid circles), neutron-deficient nuclei with $N/Z < 1$ (crosses), neutron-rich nuclei with $N/Z > 1.5$ (solid squares).}
\label{all_isotopes_T4_RQRPA_Goriely.eps}
\end{figure}

In the case of cross sections averaged over the supernova neutrino flux for ($T = 4$, $\eta = 0$), the impact of the self-consistent and complete calculations including all relevant multipoles is explored in view of previous knowledge on neutrino induced reactions on a large-scale basis within the OPb pool. Figure~\ref{all_isotopes_T4_RQRPA_Goriely.eps} shows the ratio of the cross sections averaged over supernova neutrino flux of the present work and ETFSI + CQRPA~\cite{Bor.00} model. In order to identify nuclei with pronounced discrepancies, the following groups of nuclei are separately denoted in figure: stable nuclei, proton-rich nuclei defined by $N/Z < 1$ and neutron-rich nuclei with $N/Z > 1.5$. For target nuclei where the cross sections are available for both models, the present large-scale calculations result mainly in systematically larger cross sections, for most of nuclei up to the factor of 4, and for smaller number of nuclei up to the factor of 7. However, for a limited set of medium mass nuclei the cross sections of this work are smaller than the ETFSI+CQRPA ones, and these are mainly limited to neutron deficient nuclei in the range $50 \lessapprox A \lessapprox 100$. The reason is the anomalous increase of the cross sections in the ETFSI + CQRPA model, when moving along a particular isotope chain from $N = Z$ nucleus toward proton rich nuclei (see Figs.~\ref{csm_Ni_isotopes_T4.eps} and ~\ref{csm_Sn_isotopes_T4.eps}). This behavior has not been observed in the present analysis. Apart from this anomaly, one can observe systematic linear increase of the ratio of cross sections with the nuclear mass. As already discussed in the case of isotopic chains, in addition to the self-consistent implementation of the RNEDF, the model employed in the present work includes a complete set of transition operators up to $J = 5$ multipoles which enhance the overall cross sections. As shown in Fig.~\ref{all_isotopes_T4_RQRPA_Goriely.eps} the linear trend is particularly apparent in the case of neutron-rich nuclei ($N/Z > 1.5$), where the absolute values of the cross sections are relatively large in comparison to other nuclei under consideration.

\begin{figure}
%\vspace*{1cm}
\centering
\includegraphics[scale=0.7]{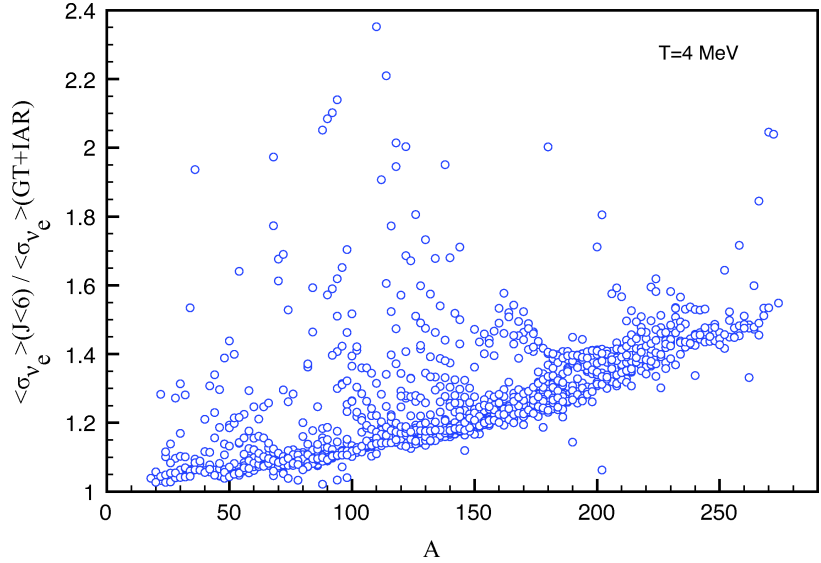}
\caption{(Color online) The ratio of the flux-averaged cross sections obtained by taking into account all multipole contributions and cross sections containing only Fermi and Gamow-Teller transitions. The Fermi-Dirac distribution was used for the neutrino spectrum with $T = 4$ MeV and $\eta = 0$.}
\label{all_isotopes_T4_all_vs_gt_iar.eps}
\end{figure}

To further explore the role of forbidden transitions in the OPb pool, in Fig.~\ref{all_isotopes_T4_all_vs_gt_iar.eps} we plot the ratio of cross sections taking into account all multipoles up to and including $J = 5$ with cross sections that only include the contributions from the Fermi and Gamow-Teller transitions. A trend very similar to the one in the previous figure appears, which indicates that the forbidden transitions contribute a larger fraction of the total cross sections in nuclei with higher mass. In heavier nuclei, and neutron-rich nuclei in particular, the difference between the proton and neutron numbers grows large, to the extent that in the very heavy nuclei neutrons occupy a full shell more above the protons. However, even in cases where there is not enough neutrons to occupy a full shell, additional neutrons move closer in energy to the orbits of the next shell, and therefore make the forbidden transitions less suppressed. The present analysis provides quantitative estimates for the mass dependence of the underlying structure in the neutrino-nucleus cross sections: in the mass region below $A \le 50$ the forbidden transitions contribute less than 10\% of the total flux-averaged cross sections, but their contribution grows practically linear with mass and can provide up to 50\% of the total cross section. 
Therefore, in order to provide a realistic description of neutrino-nucleus reactions, as well as in the corresponding astrophysical applications, models must take into account more than the simplest Fermi and Gamow-Teller terms. 

\begin{figure}
%\vspace*{1cm}
\centering
\includegraphics[scale=0.5]{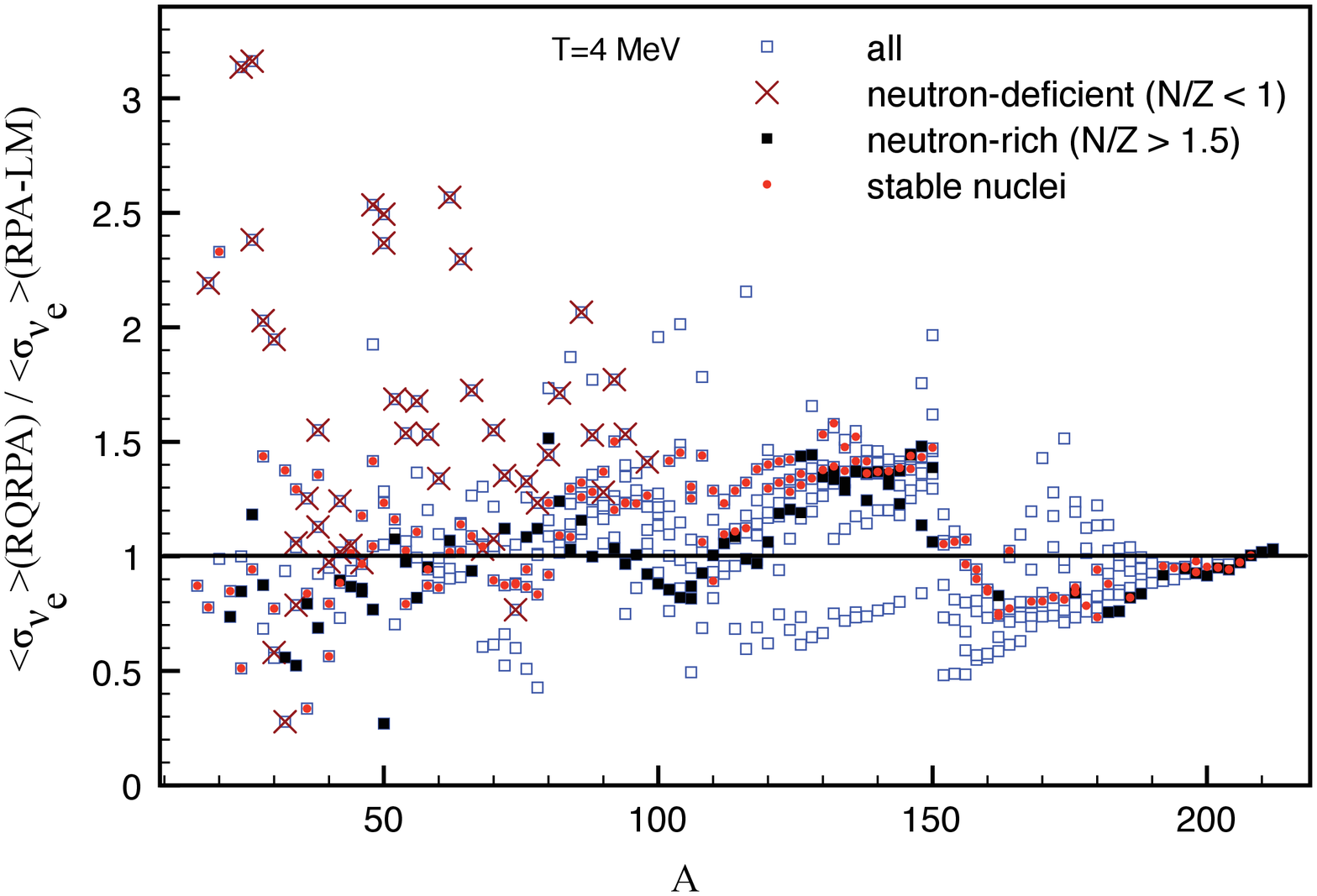}
\caption{(Color online) The ratio of the cross sections averaged over supernova neutrino distribution ($T = 4$ MeV, $\eta = 0$) calculated using the RHB+RQRPA and RPA (WS+LM)~\protect\cite{Lan.01} models. Various groups of nuclei are separately denoted: stable nuclei (solid circles), neuton-deficient nuclei with $N/Z < 1$ (crosses), neutron-rich nuclei with $N/Z > 1.5$ (solid squares).}
\label{all_isotopes_T4_RQRPA_Langanke.eps}
\end{figure}

In Fig.~\ref{all_isotopes_T4_RQRPA_Langanke.eps} the ratio of the averaged cross sections of this work and RPA (WS + LM)~\cite{Lan.01} is shown. Qualitative agreement between the two models is obtained and differences are within a factor of 2. The ratio exhibits a mild mass dependence, i.e. the model based on RQRPA provides larger cross sections mainly for nuclei up to $A \approx 150$, while for heavier systems the cross sections are up to 50\% smaller when compared to the ones previously reported with the RPA (WS + LM) model. These discrepancies occur for particular isotopic chains, however, there is no global systematic over- or underestimation of the cross-sections with respect to the RPA (WS + LM) approach, indicating that the deviations mainly originate from the different nuclear ground state and the residual interaction. The lack of a linear trend for neutron-rich nuclei 
(in comparison to Figs.~\ref{all_isotopes_T4_all_vs_gt_iar.eps} and ~\ref{all_isotopes_T4_RQRPA_Goriely.eps}), along with the fact that RPA (WS+LM) model includes all transitions up to and including $J = 3$ points to the importance of the lowest forbidden transitions in the quantitative description of semileptonic processes in nuclei. The largest deviations appear for low mass proton-rich nuclei, where the $(\nu_e,e^-)$ reactions become suppressed and the absolute values of the cross sections are actually very low, making them very sensitive to the structural details of the ground state. In comparison to the RPA (WS + LM) model, the main advantage of the present work
is implementation of fully self-consistent theory framework in description of nuclear
sector in neutrino induced reactions. In this way  predictions for the cross sections become feasible
not only in the region of stable nuclei and those that are covered by the RPA (WS+LM) model , but
also in the unknown regions of the nuclide map of relevance for the stellar processes and 
nucleosynthesis.  

% TF %%%%%%%%%%%%%%%%%%%%%%%%%%%%%%%%%%%%%%%%%%%%%%%%%%%%%%%%%%%%%%%%%%%%%%%%%%%%%%%%%%
In addition to the Fermi-Dirac spectrum Eq.~(\ref{fermidirac}), we also apply the
distribution obtained from a core-collapse supernova
simulation~\cite{Fischer2012}.
It can be described by the $\alpha$-fit as follows
\begin{equation}
f_\alpha(E_{\nu}) = \frac{1}{\Gamma(1+\alpha)} \left( \frac{1 + \alpha}{\left\langle E_{\nu} \right\rangle} \right)^{1+\alpha} E_{\nu}^{\alpha} \exp{\left( -\frac{(1+\alpha)E_{\nu}}{\left\langle E_{\nu} \right\rangle} \right)},
\label{eq:alphadist}
\end{equation}
where the parameter $\alpha$ is determined from the mean neutrino energy and the
root-mean-square of the neutrino energy as~\cite{Keil2003}
\begin{equation}
\left\langle E_{\nu}^{2} \right\rangle = \frac{\alpha + 2}{\alpha + 1} \left\langle E_{\nu} \right\rangle^{2}.
\end{equation}
These data are taken from the core-collapse supernova explosion simulations
of Ref.~\cite{Fischer:2010}.
The supernova model is based on general relativistic radiation hydrodynamics
and three-flavor Boltzmann neutrino transport in spherical symmetry.
It also employs a nuclear equation of state~\cite{Shen:1998} and a modern set
of neutrino opacities (for details, see Table~1 in Ref.~\cite{Fischer2012}
and the references therein).
The simulations were launched from massive progenitor stars, the 8.8~M$_\odot$
O-Ne-Mg-core and the more massive Fe-core progenitors of 10.8 and 18~M$_\odot$.
Note that while for the low-mass star neutrino driven explosions can be obtained
in spherically symmetric models, neutrino heating/cooling was enhanced artificially
in order to trigger the onset of explosion for the more massive stars
(for details, see Ref.~\cite{Fischer:2010}).
For the current discussions, we select the data from the 18~M$_\odot$ explosion model.
Nevertheless, all simulations exhibit the same intrinsic feature of the
continuously decreasing neutrino luminosities and average energies
(see, e.g., Figure~14 in Ref.~\cite{Fischer:2010})
after the supernova explosion has been launched.
This important aspect is related to the continuous emission of neutrinos of all flavors,
which deleptonizes the central protoneutron star on timescales of tens of seconds.
Note that at the onset of the supernova explosion, the protoneutron star is hot
(10--40~MeV) and still lepton rich, in which sense it differs from the final
supernova remnant neutron star.
In order to characterize the dynamical evolution of the neutrino spectra during the
protoneutron star deleptonization, we select three different post bounce times after
the explosion has been launched, $t_{pb} = 1$, 5, and 20~seconds
(see Table~\ref{Tab:spectra} for the corresponding neutrino energies).

\begin{figure}
\vspace*{1cm}
\centering
\includegraphics[scale=0.5]{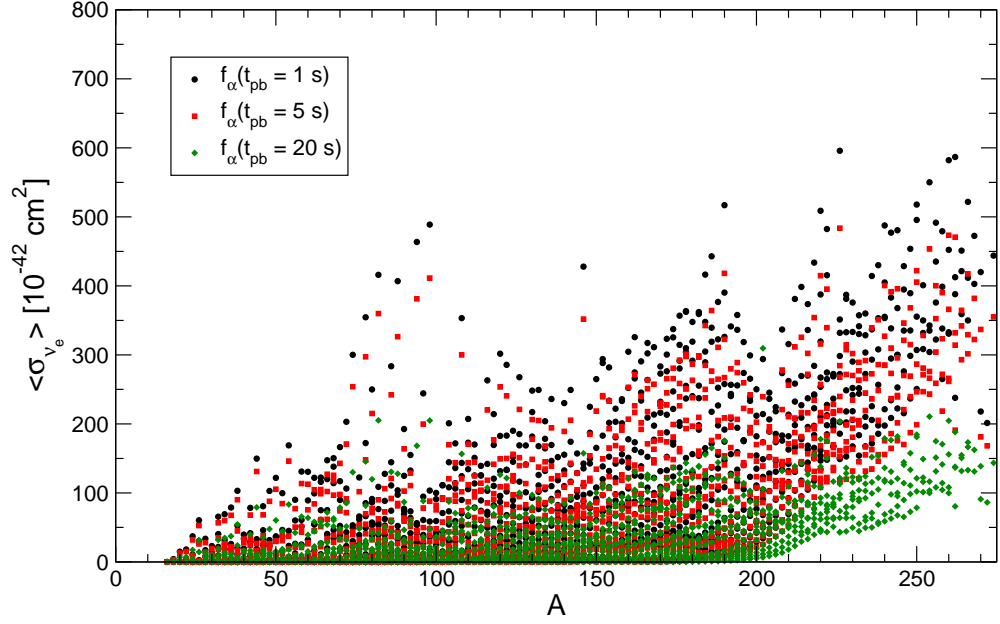}
\caption{(Color online) Flux-averaged cross sections of all nuclei in the OPb pool using the supernova simulation spectra, comparing the different post-bounce times $t_{pb} = 1$, 5, 20~s.}
\label{fig:falphacs}
\end{figure}

\begin{figure}
\vspace*{1cm}
\centering
\includegraphics[scale=0.5]{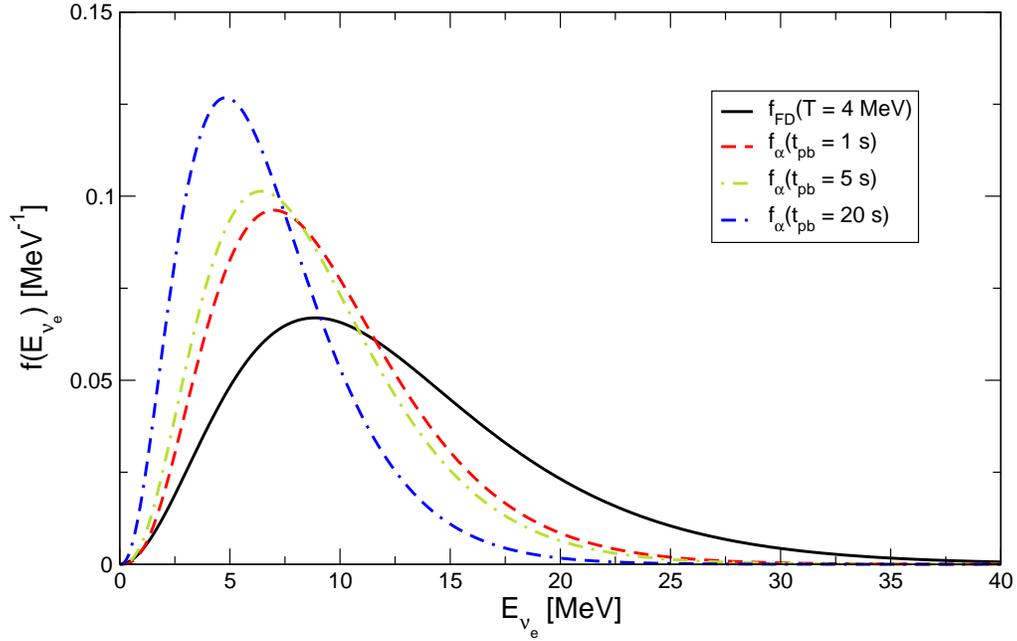}
\caption{(Color online) Fermi-Dirac $f_\text{FD}$ and supernova neutrino spectra $f_\alpha$. The latter correspond to the selected post-bounce times $t_{pb} = 1$, 5, 20~s.}
\label{fig:spectra}
\end{figure}

Once the neutrinos have decoupled from matter at the sphere of last scattering,
which depends on neutrino flavor and energy, they can still contribute to charge
and neutral current processes involving heavy nuclei at large distance
($ \sim 10^3 - 10^4$~km) relevant for the nucleosynthesis, e.g.,
the $\nu p$ process~\cite{Froehlich:2006} and the $r$ process~\cite{Qian:2007}.
At conditions where these processes can occur, the zero-temperature approach employed
in the current paper for the calculation of the cross sections is a valid approximation.
The protoneutron star deleptonization with the continuously reducing average
neutrino energies has important consequences for the neutrino flux-averaged cross
sections, as shown in Fig.~\ref{fig:falphacs}.
In comparison to the Fermi-Dirac spectrum with $T = 4$ MeV shown in
Fig.~\ref{all_isotopes_T4.eps}, the supernova simulation spectrum results in significantly
lower flux-averaged cross sections by more than a factor of two for all nuclei
under investigation, even shortly ($t_{pb} = 1$~s) after the onset of explosion.
Note that the supernova neutrino temperature, which can be defined as the matter
temperature at the neutrinosphere, is slightly higher with $T \simeq 4.5-5$~MeV than the
assumed $T = 4$~MeV of the Fermi-Dirac spectrum.
However, the Fermi-Dirac neutrino spectrum is significantly broader than those taken from the
supernova simulations, which can be seen by comparing mean neutrino energies
$\langle E \rangle$ and the mean-square energies $\langle E^2 \rangle$ of $f_\text{FD}$
and $f_\alpha$.
For comparison, we also plot the spectra in Fig.~\ref{fig:spectra}.
The above mentioned difference between Fermi-Dirac spectra and supernova simulation spectra
even increases during the ongoing explosion at late times, i.e. the supernova
spectra shift increasingly towards Maxwell-Boltzmann like spectra.
These findings are summarized in Table~\ref{Tab:spectra}, comparing the
Fermi-Dirac spectra ($f_\text{FD}$) and the supernova simulation spectra characterized
by the $\alpha$-fit ($f_\alpha$) at different post-bounce times after the
explosion has been launched.
\begin{table*}[htp]
\centering
\caption{Neutrino mean and mean-square energies for the different neutrino distributions.}
\begin{tabular}{ c c c c}
\hline
\hline
 & $t_{pb}$\footnote{Post bounce simulation time for the supernova spectra} $[s]$ & $\langle E \rangle$ [MeV] & $\langle E^2 \rangle$ [MeV$^2$] \\
\hline
$f_\text{FD}$ & & 12.60 & 207.03 \\
$f_\alpha$ & 1 & 9.32 & 108.57 \\
$f_\alpha$ & 5 & 8.68 & 95.29 \\
$f_\alpha$ & 20 & 6.73 & 58.32 \\
\hline
\end{tabular}
\label{Tab:spectra}
\end{table*}

\section{Conclusion}

We have employed a fully self-consistent model based on the RNEDF in large-scale calculations of charged-current neutrino-nucleus cross sections in the OPb pool of target nuclei, spanning the range $8 \le Z \le 82$ and $8 \le N \le 182$. The two main advantages of this approach are: (i) self-consistent modeling of all relevant transition matrix elements involving open-shell nuclei, without any additional adjustments of the model parameters to the nuclear target under consideration, and (ii) treating the allowed and the forbidden transitions on an equal footing, i.e., transitions of all relevant multipoles
are taken into account. For the purpose of the present work, advanced computational framework has been developed, based on a parallel computing scheme using the Message Passing Interface for the implementation on cluster and grid computer systems. Reasonable agreement of the cross sections with the only available experimental data, for $^{12}$C and $^{56}$Fe, support the course of systematic calculations throughout the nuclide map. The results include a complete set of the cross sections for the OPb pool, calculated for the range of neutrino energies $E_{\nu_e} = 0 - 100$ MeV. The cross sections were averaged over the experimental Michel spectrum, the neutrino spectrum described by the Fermi-Dirac distribution for the range of temperatures $0 \le T \le 10$ MeV and chemical potential $\eta = 0$. In addition, the averaged cross sections have been calculated using $\alpha$-fit neutrino fluxes, taken from a recent core-collapse supernova explosion model that is based on three-flavor Boltzmann neutrino transport. Comparing Fermi-Dirac and supernova simulation neutrino spectra, we found that the latter result in significantly smaller flux-averaged cross sections due to the more pinched supernova neutrino spectra. This aspect even increases during the ongoing deleptonization on the order of 10--20~s after which the cross sections become negligible and charged-current processes with nuclei have ceased.

Note that the neutrino spectra form the supernova simulations do not take into account possible collective neutrino-flavor oscillations. These phenomena can take place after neutrino decoupling from matter and result in complete spectral swaps above a certain neutrino energy for inverted neutrino-mass hierarchy~\cite{Dasgupta:2009mg}. Because heavy-lepton flavor neutrinos have higher average energies, being less strongly bound to matter in the absence of charged-current weak processes, it enhances the high-energy tail of the electron-flavor neutrino spectra. This in turn may impact neutrino-induced nucleosynthesis of heavy elements at large distance~\cite{Martinez:11,Duan:11}.

Furthermore, it has been realized that medium modifications for the charged-current weak processes with neutrons and protons must be taken into account when computing neutrino transport and neutrino decoupling from matter at high densities using a nuclear equation of state~\cite{Reddy:98}. These modifications have been explored at the mean-field level and shown to increase spectral differences between electron neutrinos and antineutrinos~\cite{Martinez:12, Roberts:12, Horowitz:12}. Improved supernova simulations that take these effects into account may ultimately alter the results discussed in the current paper.

Several key features of the neutrino-nucleus cross sections have been illustrated for various supernova neutrino fluxes, indicating systematic increase of the averaged cross sections with increase of the number of neutrons in target nuclei as well as with temperature of the neutrino distribution. When going away from the valley of stability toward neutron-rich nuclei, the cross sections become considerably enhanced, while on the proton-rich side they are suppressed due to the blocking of orbitals available
for neutrino induced transitions. 

The cross sections from the present analysis have been discussed in view of previously reported large scale calculations. Current results are consistently higher than the results obtained using the ETFSI + CQRPA model which includes only the Fermi and the Gamow-Teller transitions. For lighter nuclei the RQRPA cross sections are approximately 50\% larger, while for the heaviest nuclei studied they are up to 3 times larger. This enhancement originates in part from the forbidden transitions that are fully taken into account in the present work. 
We provide important quantitative insight into the underlying structure of the neutrino-nucleus cross sections throughout the nuclide map: in the mass region below $A \le 50$ the forbidden transitions contribute less than 10\% of the total flux-averaged cross sections, but their contribution increases with mass and can provide up to 50\% of the total cross section.
For the limited set of nuclei, the differences between the present results and those of the RPA (LM + WS) model are constrained within a factor of three. There is no general trend in the cross sections ratio between the two models, and the differences arise mainly due to the different approach in the calculations of the nuclear ground state and the residual interaction in (Q)RPA.

We have shown that the self-consistent calculations based on the RNEDF, which take into account transitions of all relevant multipoles, result in differences in comparison to previously reported cross sections, and for a considerable number of target nuclei the cross sections are larger. Therefore, revised calculations in modeling stellar evolution and r-process nucleosynthesis based on self-consistent descriptions of neutrino-induced reactions may allow an updated insight into the origin of elements in the Universe. On the other hand, by employing various theory frameworks for neutrino-nucleus cross sections, one could estimate the uncertainties in the calculated element abundance patterns. The tables with the cross sections vs. neutrino energies of this work are available from the authors on request.
%
%=========================================================================================
\leftline{\bf ACKNOWLEDGMENTS}
\noindent
We gratefully acknowledge S. Goriely for providing data for the ETFSI+CQRPA calculations from Ref.~\cite{Bor.00}. This work is supported by MZOS - project 1191005-1010, the Croatian Science Foundation and by the Helmholtz International Center for FAIR within the framework of the LOEWE program launched by the State of Hesse. The computational resources have been provided in part by the Croatian National Grid Infrastructure (CRO-NGI).
T.F. is supported by the Swiss National Science Foundation under project~no.~PBBSP2-133378.
%========================================================================================

\newpage

%-----------------------------------------------------------------------
%-----------------------------------------------------------------------

%-----------------------------------------------------------------------
%-----------------------------------------------------------------------

%-----------------------------------------------------------------------

%-----------------------------------------------------------------------

%-----------------------------------------------------------------------

%-----------------------------------------------------------------------

%-----------------------------------------------------------------------

%-----------------------------------------------------------------------

%-----------------------------------------------------------------------

%-----------------------------------------------------------------------

%-----------------------------------------------------------------------

\end{document}